\documentclass[pdflatex,sn-mathphys-num]{sn-jnl}

\usepackage{natbib,aas_macros,amsmath}
\usepackage{graphicx}%
\usepackage{multirow}%
\usepackage{amsmath,amssymb,amsfonts}%
\usepackage{amsthm}%
\usepackage{mathrsfs}%
\usepackage[title]{appendix}%
\usepackage{xcolor}%
\usepackage{textcomp}%
\usepackage{manyfoot}%
\usepackage{booktabs}%
\usepackage{algorithm}%
\usepackage{algorithmicx}%
\usepackage{algpseudocode}%
\usepackage{listings}%

\theoremstyle{thmstyleone}%
\theoremstyle{thmstyletwo}%

\theoremstyle{thmstylethree}%

\newcommand{\arcsec}{$^{\prime\prime}$}
\newcommand{\carcsec}{$\!\!^{\prime\prime}$}
\newcommand{\m}[1]{\mathrm{#1}}

\newcommand{\Oiii}{{\sc[O\,iii]}}
\newcommand{\Oii}{{\sc[O\,ii]}}

\setcitestyle{super}
\setcitestyle{open={},close={}}

\begin{document}

\title[Article Title]{A UV-Luminous Galaxy at $z=11$ with Surprisingly Weak Star Formation Activity}


\author*[1]{\fnm{Yuichi} \sur{Harikane}}\email{hari@icrr.u-tokyo.ac.jp}

\author[2]{\fnm{Pablo G.} \sur{P\'erez-Gonz\'alez}}

\author[2]{\fnm{Javier} \sur{\'Alvarez-M\'arquez}}

\author[1,3,4,5]{\fnm{Masami} \sur{Ouchi}}

\author[6]{\fnm{Yurina} \sur{Nakazato}}

\author[1]{\fnm{Yoshiaki} \sur{Ono}}

\author[7]{\fnm{Kimihiko} \sur{Nakajima}}

\author[1,8]{\fnm{Hiroya} \sur{Umeda}}

\author[9,10,11]{\fnm{Yuki} \sur{Isobe}}

\author[12,3]{\fnm{Yi} \sur{Xu}}

\author[13]{\fnm{Yechi} \sur{Zhang}}

\affil[1]{\orgdiv{Institute for Cosmic Ray Research}, \orgname{The University of Tokyo}, \orgaddress{\street{5-1-5 Kashiwanoha}, \city{Kashiwa}, \postcode{277-8582}, \state{Chiba}, \country{Japan}}}
\affil[2]{\orgdiv{Centro de Astrobiolog\'{\i}a (CAB)}, \orgname{CSIC-INTA}, \orgaddress{\street{Ctra. de Ajalvir km 4}, \city{Torrej\'on de Ardoz}, \postcode{E-28850}, \state{Madrid}, \country{Spain}}}
\affil[3]{National Astronomical Observatory of Japan, 2-21-1 Osawa, Mitaka, Tokyo 181-8588, Japan}
\affil[4]{Department of Astronomical Science, SOKENDAI (The Graduate University for Advanced Studies), Osawa 2-21-1, Mitaka, Tokyo 181-8588, Japan}
\affil[5]{Kavli Institute for the Physics and Mathematics of the Universe (WPI), University of Tokyo, Kashiwa, Chiba 277-8583, Japan}
\affil[6]{\orgdiv{Center for Computational Astrophysics}, \orgname{Flatiron Institute}, \orgaddress{\street{162 5th Avenue}, \city{New York}, \postcode{10010}, \state{NY}, \country{USA}}}
\affil[7]{Institute of Liberal Arts and Science, Kanazawa University, Kakuma-machi, Kanazawa, Ishikawa 920-1192, Japan}
\affil[8]{Department of Physics, Graduate School of Science, The University of Tokyo, 7-3-1 Hongo, Bunkyo, Tokyo 113-0033, Japan}
\affil[9]{\orgdiv{Kavli Institute for Cosmology}, \orgname{University of Cambridge}, \orgaddress{\street{Madingley Road}, \city{Cambridge}, \postcode{CB3 0HA}, \state{Cambridgeshire}, \country{UK}}}
\affil[10]{\orgdiv{Cavendish Laboratory}, \orgname{University of Cambridge}, \orgaddress{\street{19 JJ Thomson Avenue}, \city{Cambridge}, \postcode{CB3 0HE}, \state{Cambridgeshire}, \country{UK}}}
\affil[11]{\orgdiv{Waseda Research Institute for Science and Engineering, Faculty of Science and Engineering}, \orgname{Waseda University}, \orgaddress{\street{3-4-1, Okubo}, \city{Shinjuku}, \postcode{169-8555}, \state{Tokyo}, \country{Japan}}}
\affil[12]{\orgdiv{Cosmic Dawn Center (DAWN)}, \orgname{Niels Bohr Institute, University of Copenhagen}, \orgaddress{\street{Jagtvej 128}, \city{Copenhagen N}, \postcode{DK-2200}, \country{Denmark}}}
\affil[13]{IPAC, California Institute of Technology, MC 314-6, 1200 E. California Boulevard, Pasadena, CA 91125, USA}






\abstract{{\bf
One of the major discoveries by the James Webb Space Telescope (JWST) is the identification of a large population of luminous galaxies at $z>10$, challenging theoretical models for early galaxy formation \cite{2022ApJ...940L..14N,2022ApJ...938L..15C,2023MNRAS.518.4755A,2023MNRAS.518.6011D,2023ApJS..265....5H,2023NatAs...7..622C,2023A&A...677A..88B,2023ApJ...946L..13F,2024ApJ...960...56H,2023Natur.622..707A,2024Natur.633..318C,2025ApJ...980..138H,2025arXiv250511263N,2023ApJ...951L...1P,2025ApJ...991..179P}. 
The unexpectedly high number density of these systems has triggered intense debate about potential differences in the physical properties of galaxies at such extreme redshifts and those at lower redshift \cite{2023MNRAS.523.3201D,2023MNRAS.522.3986F,2023MNRAS.525.3254S,2023ApJ...955L..35S,2023MNRAS.526.2665S,2024MNRAS.527.5929Y,2024JCAP...08..025H,2026arXiv260107912M}. 
However, progress has been limited by the lack of rest-frame optical diagnostics, which are critical for constraining the key properties. 
Here we present deep JWST/MIRI observations of a UV-luminous galaxy at $z=11.04$, CEERS2-588, only 400\,Myr after the Big Bang. 
CEERS2-588 is detected in the MIRI F560W and F770W bands, while deep MIRI/MRS spectroscopy yields no detection of H$\alpha$ or [OIII]$\lambda5007$ line, revealing a prominent Balmer break detected for the first time at $z>10$.
Spectral energy distribution (SED) fitting indicates an extended star formation history possibly reaching $z\gtrsim15$, followed by rapid quenching within the recent $\sim10$ Myr, in stark contrast to other $z>10$ galaxies.
The MIRI detections also significantly improve our stellar mass estimate to $\mathrm{log}(M_*/M_\odot)=9.1^{+0.1}_{-0.1}$, making CEERS2-588 the most massive galaxy securely confirmed at $z>10$. 
Remarkably, the inferred gas-phase metallicity is near solar, exceeding predictions from current theoretical models. 
These results suggest that efficient starbursts play a key role in producing the abundant luminous galaxy population in the early universe.
}
}

\maketitle

The target galaxy CEERS2-588 (R.A.$=14{:}19{:}37.59$, Decl.$=+52{:}56{:}43.8$) was first identified in NIRCam imaging of the CEERS field \cite{2023ApJ...946L..13F}.
Its spectroscopic redshift, $z_{\mathrm{spec}}=11.04$, was determined from the detections of the Lyman break and the [O\,\textsc{ii}]$\lambda3727$ emission line in the NIRSpec PRISM spectrum \cite{2024ApJ...960...56H} (Extended Data Fig.~\ref{fig_spec2}).
With an absolute ultraviolet magnitude of $M_{\mathrm{UV}}=-20.4$ mag (Extended Data Table~\ref{tab_target}), CEERS2-588 is among the most UV-luminous galaxies known at $z>11$, a population whose abundance was not anticipated by pre-JWST theoretical models \cite{2024ApJ...960...56H}.
Its extended morphology, with an effective radius of $r_{\mathrm{e}}\sim450$ pc (Extended Data Table~\ref{tab_target}), together with the non-detection of high-ionization emission lines in the NIRSpec spectrum, indicates the absence of strong active galactic nucleus (AGN) activity in CEERS2-588.
The galaxy is therefore classified as a typical extended system \cite{2025ApJ...980..138H}, similar to JADES-GS-z14-0 at $z=14.2$ \cite{2024Natur.633..318C}.
Despite its extreme luminosity, previous estimates of the stellar mass of CEERS2-588 were highly uncertain, $8.1\lesssim \log(M_*/M_\odot)\lesssim9.5$ \cite{2024ApJ...960...56H,2023Natur.622..707A}, because existing NIRCam and NIRSpec observations probed only rest-frame ultraviolet wavelengths.

\begin{figure}[t]
\centering
\includegraphics[width=0.99\textwidth, bb=11 4 700 428]{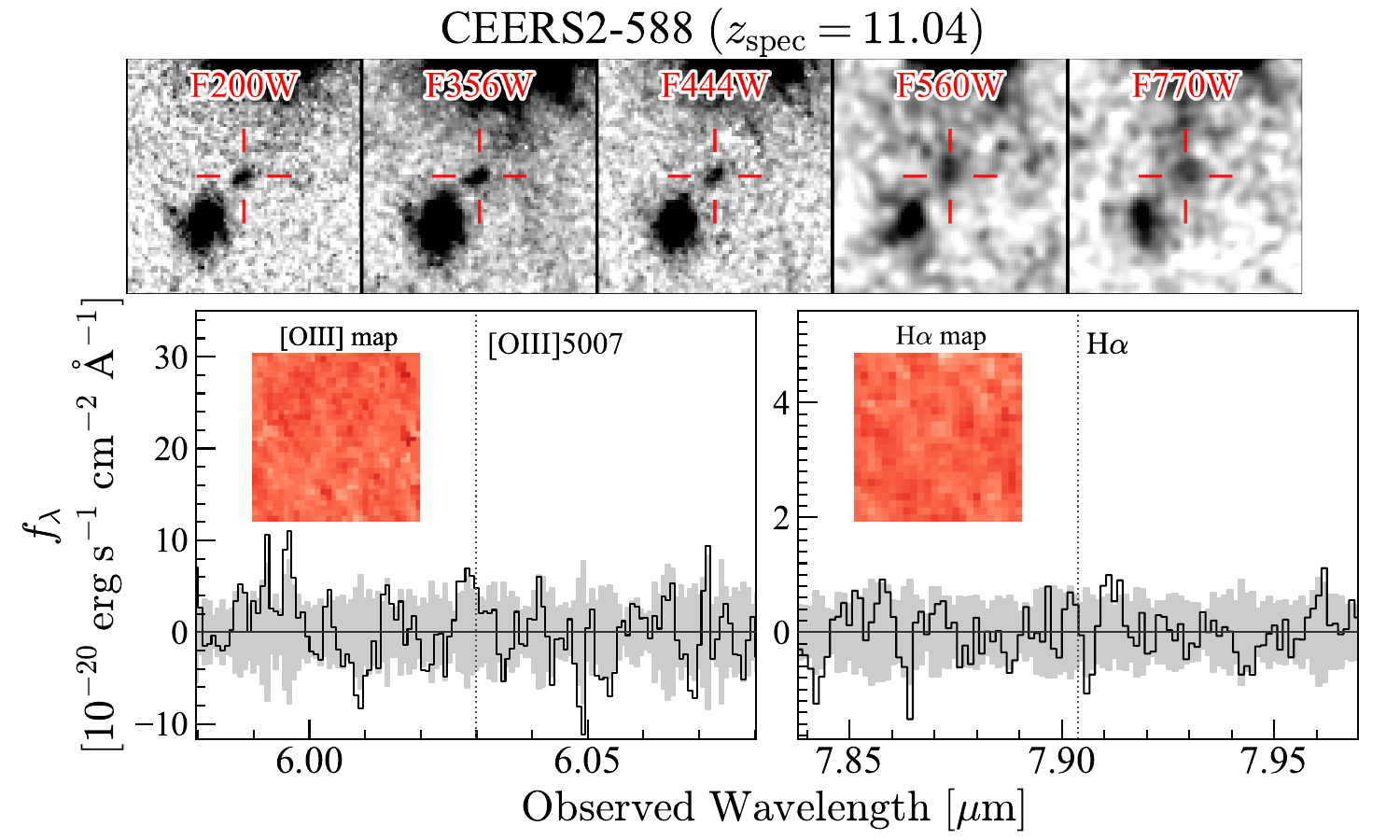}
\caption{{\bf JWST MIRI observations of CEERS2-588.}
Top panels show 6\arcsec$\times$6\arcsec cutout images obtained with JWST NIRCam and MIRI, with $\sim6\sigma$ detections in the MIRI F560W and F770W bands.
The bottom panels present the extracted one-dimensional MIRI/MRS spectra (black), with $1\sigma$ uncertainties indicated in grey, around the expected wavelengths of the [O\,\textsc{iii}]$\lambda5007$ and H$\alpha$ lines.
Inset panels show the MIRI/MRS maps at these wavelengths, integrated over $\pm100$ km s$^{-1}$.
No significant [O\,\textsc{iii}]$\lambda5007$ or H$\alpha$ emission is detected in the spectra, indicating that the MIRI F560W and F770W detections are dominated by rest-frame optical stellar continuum emission.
}
\label{fig_spec}
\end{figure}

In JWST Cycle~3, CEERS2-588 was observed with deep MIRI \cite{2023PASP..135d8003W} imaging and spectroscopy (Methods).
Figure~\ref{fig_spec} presents the MIRI data obtained in our observations.
The galaxy is robustly detected in the MIRI F560W and F770W bands.
However, despite long total integration times of 9.5 and 17.8 hours, neither [O\,\textsc{iii}]$\lambda5007$ nor H$\alpha$ is significantly detected in the MIRI/MRS spectra, with $3\sigma$ upper limits of $<2.4\times10^{-18}$ and $<5.1\times10^{-19}\,\mathrm{erg\,s^{-1}\,cm^{-2}}$, respectively.
The inferred rest-frame equivalent widths are constrained to be $\lesssim200-400$~\AA\ (Extended Data Table~\ref{tab_spec}), substantially lower than those measured for comparably luminous galaxies at high redshift \cite{2024MNRAS.533.1111E}, implying a stellar age of $\gtrsim20$ Myr for an instantaneous burst, or $\gtrsim100$ Myr for a constant star formation history \cite{2025A&A...701A..31P,2011MNRAS.415.2920I}.

\begin{figure}[t]
\centering
\includegraphics[width=0.99\textwidth,bb=10 8 935 318]{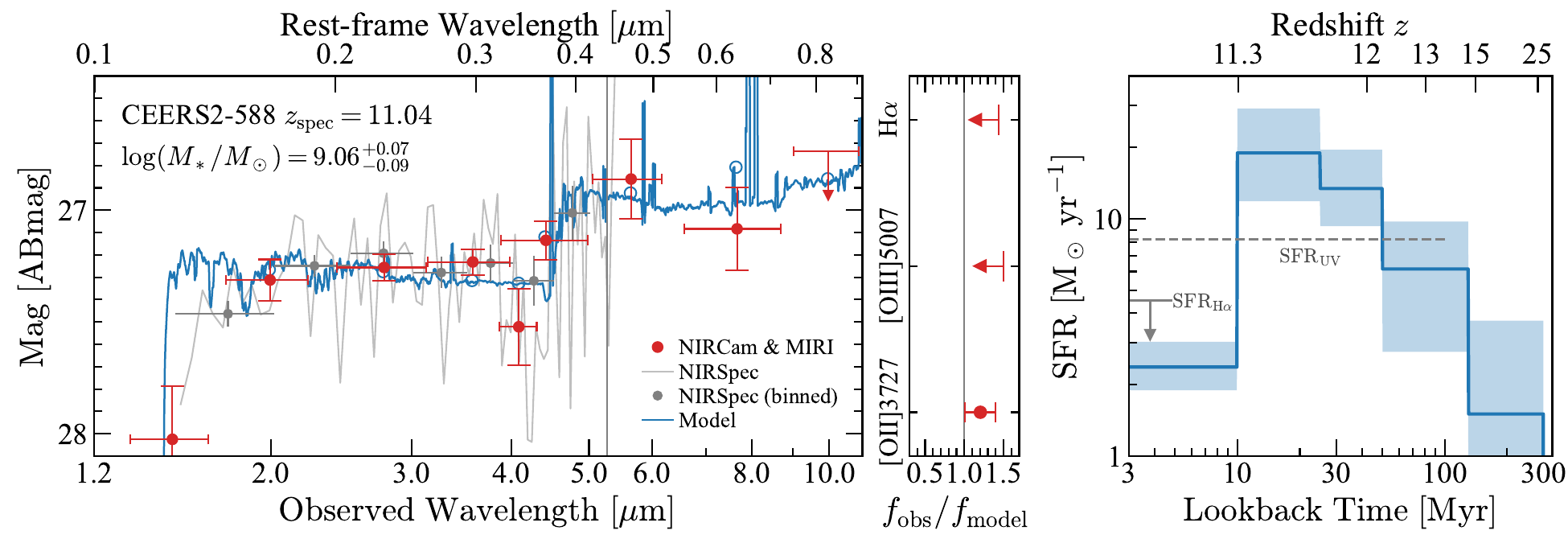}
\caption{\textbf{Results of SED fitting for CEERS2-588.}
Left panel shows the SED. 
Red circles indicate the observed photometric measurements, while the grey line and filled circles show the smoothed NIRSpec spectrum and its binned constraints. 
Both the NIRCam and MIRI photometry and NIRSpec spectroscopy indicates the existence of a prominent Balmer break.
The blue line and open circles represent the best-fitting model. 
Middle panel compares the observed and model-predicted emission-line fluxes for H$\alpha$, \Oiii$\lambda5007$, and \Oii$\lambda3727$, demonstrating good agreement between the model and the data. 
Right panel presents the inferred star formation history. 
CEERS2-588 undergoes rapid quenching within the recent $\sim10$ Myr, consistent with star formation rates inferred from H$\alpha$ (grey solid line) and UV (grey dashed line) emission, which probe characteristic timescales of $\sim5$ Myr and $\sim100$ Myr, respectively.
}
\label{fig_sed}
\end{figure}

Given the low equivalent widths of H$\alpha$ and {\Oiii}, the MIRI broadband fluxes are not dominated by emission-line contributions. 
Instead, they trace the rest-frame optical stellar continuum. 
Fig. \ref{fig_sed} shows the SED of CEERS2-588. 
The MIRI F560W and F770W magnitudes tracing the rest-frame wavelengths of $>4000$ \AA\ are brighter than the shorter-wavelength NIRCam measurements by $\sim0.3$ mag, revealing a prominent Balmer break. 
This represents the first clear detection of a Balmer break in a galaxy at $z>10$. 
The NIRSpec spectrum independently supports the presence of this Balmer break. 
SED fitting (Methods) indicates that star formation began at $z \sim 15-25$, corresponding to $\sim100-300$ Myr after the Big Bang, followed by a sharp decline in the star formation rate within the recent $\sim10$ Myr. 
This star formation history is consistent with star formation rates inferred from the UV continuum and the H$\alpha$ upper limits (Methods).

\begin{figure}[t]
\centering
\includegraphics[width=0.46\textwidth,bb=3 20 393 350]{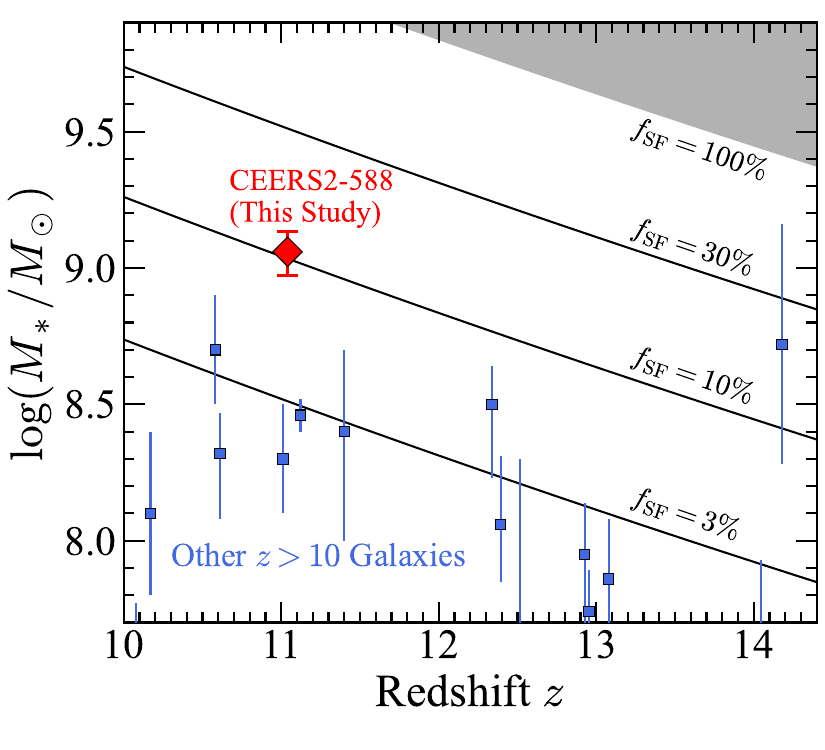}
\includegraphics[width=0.53\textwidth,bb=4 4 430 315]{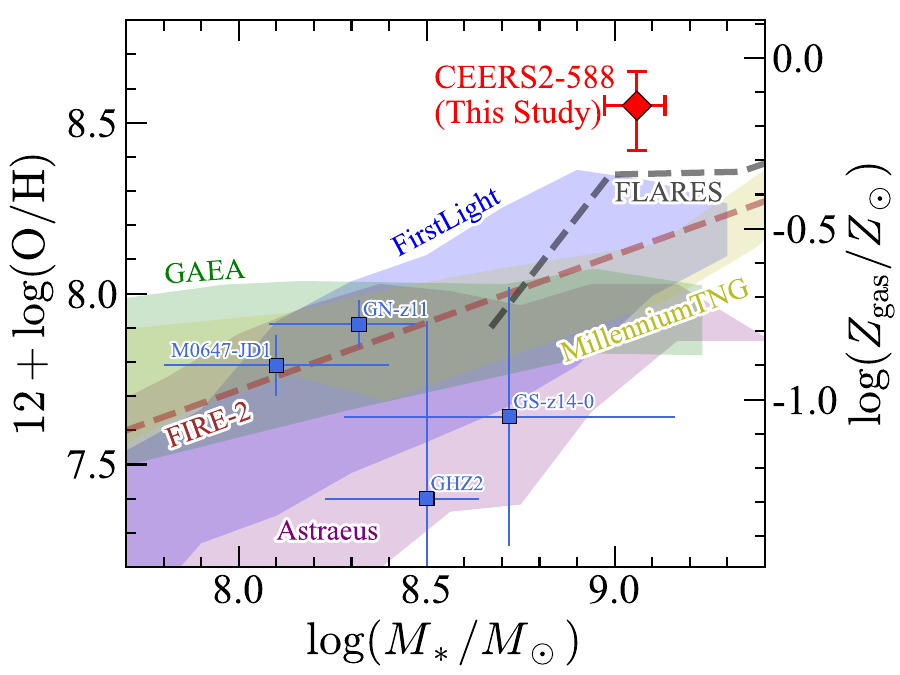}
\caption{\textbf{Basic properties of CEERS2-588 in comparison with other galaxies and theoretical models.}
Left panel shows stellar mass as a function of redshift. 
CEERS2-588 (red diamond) is the most massive galaxy known at $z>10$ among spectroscopically confirmed galaxies (blue squares) \cite{2025arXiv250511263N,2025NatAs...9..729H,2024Natur.633..318C,2025Natur.639..897W,2023NatAs...7..622C,2023ApJ...957L..34W,2025arXiv250722888W,2025arXiv251103035C,2023Natur.622..707A,2025ApJ...988L..10K,2025arXiv251202997C,2024ApJ...973....8H,2025NatAs...9..155Z}. 
Black curves indicate the stellar mass expected for the most massive dark matter halo within the CEERS survey volume at each redshift ($\m{log}(M_\m{h}/M_\odot)=10.8$ at $z=11$) for different assumed integrated star formation efficiencies, while the grey shaded region marks values forbidden by the $\Lambda$CDM cosmology. 
These curves are shown to illustrate the relative extremeness of CEERS2-588 within the CEERS volume, while the comparison galaxies are not restricted to the CEERS field.
Right panel shows the mass-metallicity relation. 
CEERS2-588 is the most metal-rich galaxy identified at $z>10$, with a metallicity $\sim5-10$ times higher than that of other galaxies with metallicity measurements at similar redshifts \cite{2025arXiv251219695H,2024ApJ...975..245C,2025A&A...695A.250A,2024ApJ...973...81H,2025NatAs...9..155Z}. 
Its metallicity also exceeds most theoretical predictions at $z \sim 11$ (Methods), indicating rapid metal enrichment in the early Universe.
}
\label{fig_Ms}
\end{figure}

CEERS2-588 exhibits properties that are exceptional among galaxies at $z>10$ (Fig. \ref{fig_Ms}) \cite{2025arXiv250511263N,2025NatAs...9..729H,2024Natur.633..318C,2025Natur.639..897W,2023NatAs...7..622C,2023ApJ...957L..34W,2025arXiv250722888W,2025arXiv251103035C,2023Natur.622..707A,2025ApJ...988L..10K,2025arXiv251202997C,2024ApJ...973....8H,2024ApJ...972..143C}. 
The measured stellar mass, $\log(M_*/M_\odot) \simeq 9.1$, makes CEERS2-588 the most massive galaxy known at $z>10$ without evidence for AGN activity. 
From the detection of {\Oii} and the non-detection of H$\alpha$ and {\Oiii}, we constrain a gas-phase metallicity close to solar, $12+\log(\mathrm{O/H}) \simeq 8.6$, using multiple strong-line diagnostics (Methods). 
Such a massive and metal-rich system has not previously been reported at $z>10$ and is not predicted by current theoretical models.

\begin{figure}[t]
\centering
\includegraphics[width=0.48\textwidth,bb=3 13 350 350]{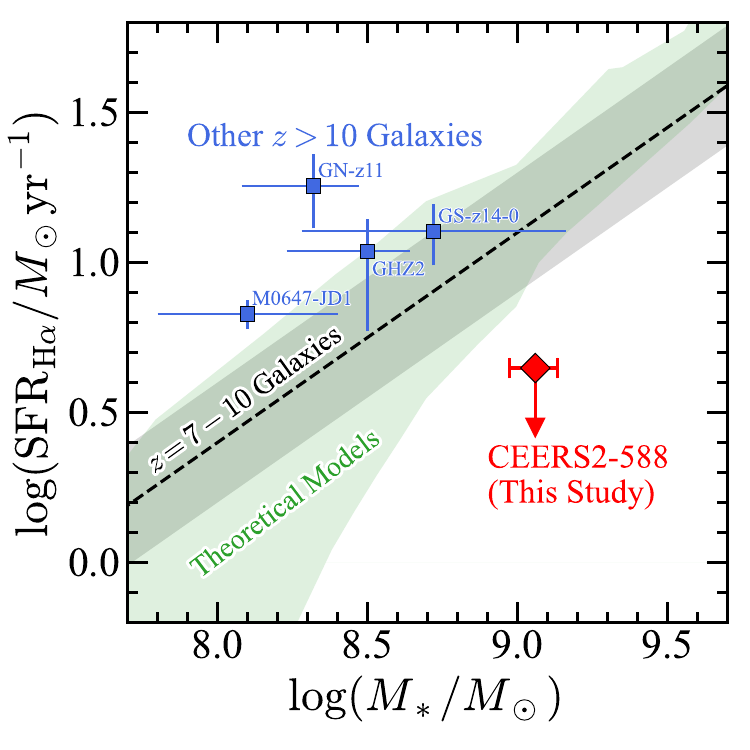}
\includegraphics[width=0.50\textwidth,bb=3 31 350 356]{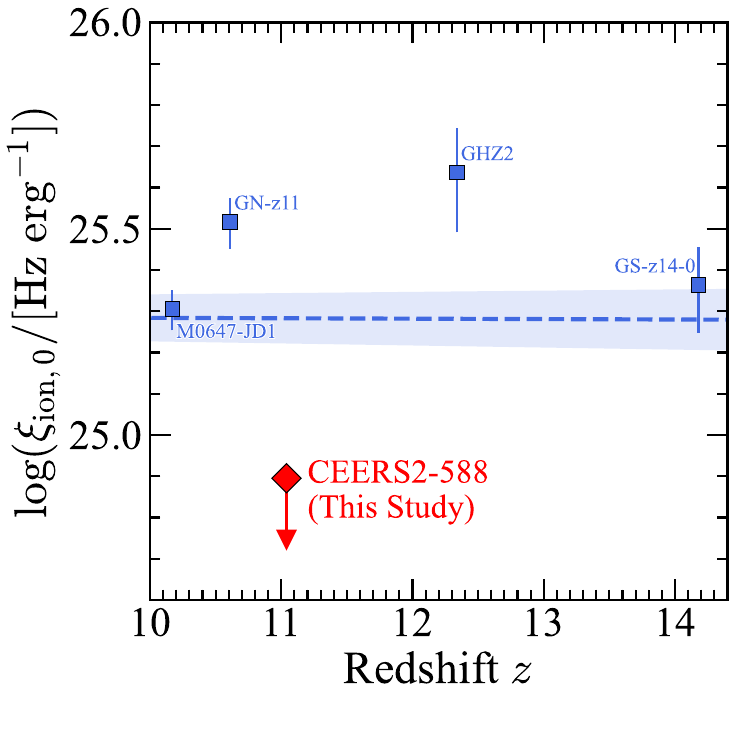}
\caption{\textbf{H$\alpha$ deficiency in CEERS2-588.}
Left panel shows the star formation main sequence, with star formation rates derived from H$\alpha$ emission. 
CEERS2-588 (red diamond; $3\sigma$ upper limit) exhibits a remarkably lower H$\alpha$-based star formation rate than other galaxies at $z>10$ \cite{2025arXiv251219695H,2024ApJ...975..245C,2025A&A...695A.250A,2024ApJ...973...81H}. 
The black dashed line and shaded region indicate the star formation main sequence at $z \sim 7-10$ and its $1\sigma$ scatter, while the green shaded region denotes the range of theoretical predictions at $z \sim 11$ (Methods). 
Right panel shows the ionizing photon production efficiency, $\xi_{\mathrm{ion}}$, for galaxies at $z>10$. 
The measured $\xi_{\mathrm{ion}}$ of CEERS2-588 (red diamond) is significantly lower than that of other galaxies at similar redshifts (blue squares) \cite{2025arXiv251219695H,2024ApJ...975..245C,2025A&A...695A.250A,2024ApJ...973...81H,2025NatAs...9..155Z}.
The blue dashed line represents an extrapolation of the empirical relation \cite{2024MNRAS.535.2998S}.
}
\label{fig_SFR}
\end{figure}

The weak H$\alpha$ emission contrasts sharply with that of other galaxies at $z>10$ (Fig. \ref{fig_SFR}). 
Despite its large stellar mass, the H$\alpha$-based star formation rate lies well below those of other galaxies at similar redshifts and falls significantly beneath the star formation rate-stellar mass relations established at $z=7-10$. 
The ionizing photon production efficiency, inferred from the ratio of H$\alpha$ to UV luminosity, is constrained to $\log(\xi_{\mathrm{ion}}/\mathrm{[Hz\,erg^{-1}]}) < 24.9$ ($3\sigma$), again lower than typical values at these redshifts.

Together, these results indicate that CEERS2-588 experienced a period of efficient starburst activity in the past, followed by recent quenching.
This behavior contrasts with other previously reported galaxies at $z>10$, which typically exhibit strong rest-frame optical emission lines consistent with ongoing starburst activity \cite{2023A&A...677A..88B,2025ApJ...988L..10K,2024ApJ...975..245C,2025A&A...695A.250A,2024ApJ...973...81H,2025arXiv251219695H}.
CEERS2-588 therefore represents the first identification of a galaxy in a post-starburst (or ``mini-quenching") phase at $z>10$.

The star formation history inferred for CEERS2-588 is consistent with a scenario in which star formation in early galaxies is highly bursty, with frequent episodes of intense star formation driving their bright UV luminosities \cite{2023MNRAS.525.3254S,2023ApJ...955L..35S,2023MNRAS.526.2665S,2026arXiv260107912M}.
However, the stellar mass of CEERS2-588 implies an integrated star formation efficiency of $\sim10\%$, exceeding the predictions of theoretical models published prior to the JWST launch ($\sim3-5\%$ for halos with $\log(M_{\rm h}/M_\odot)\simeq10.8$ at $z\gtrsim10$) \cite{2018MNRAS.477.1822M,2020MNRAS.499.5702B}.
The resulting stellar mass surface density of $\sim1700~M_\odot~\mathrm{pc^{-2}}$ is comparable to that of local globular clusters and elliptical galaxies \cite{2014MNRAS.443.1151N}, systems that are also thought to have formed a large fraction of their stars efficiently at early times.
These results indicate that high star formation efficiency contributes significantly to the UV-bright nature of CEERS2-588.

The SED and inferred star formation history of CEERS2-588 resemble that of recently quenched galaxies at $z<10$ \cite{2024Natur.629...53L,2025A&A...697A..88L}, but at a substantially higher stellar mass.
Cosmological simulations at $5<z<15$ predict star formation histories characterized by recurrent, intense bursts interspersed with mini-quiescent phases lasting $\sim10-100$ Myr, during which galaxies can temporarily exhibit suppressed nebular emission while retaining substantial stellar masses \cite{2018MNRAS.480.4842C,2026arXiv260107912M,2025ApJ...991L...4M}.
The combination of strong UV continuum and weak H$\alpha$ emission in CEERS2-588 is naturally explained if the galaxy is observed during such a post-burst lull, providing empirical constraints on the frequency and duration of quenching episodes at $z>10$.
However, the inability of current models to simultaneously reproduce both the extreme stellar mass and the strongly suppressed H$\alpha$ emission suggests that quenching mechanisms beyond stellar feedback may play an important role in massive galaxy formation at $z>10$, such as AGN feedback \cite{2025A&A...704A.248Q} and/or radiation-driven outflows \cite{2023MNRAS.522.3986F}.
These results reveal that massive galaxies in the first few hundred million years of cosmic history experienced star formation that was both more efficient and more rapidly quenched than predicted by theoretical models.

\clearpage

\section*{Methods}\label{ss_method}

\newcounter{extfigure}
\renewcommand{\thefigure}{\arabic{extfigure}}
\renewcommand{\figurename}{Extended Data Fig.}
\renewcommand{\theHfigure}{ext.\arabic{extfigure}}
\setcounter{figure}{0}

\subsection*{Adopted cosmology and conventions}

Throughout this work, we adopted the \textit{Planck} cosmological parameters from the TT, TE, EE$+$lowP$+$lensing$+$BAO analysis \cite{2020A&A...641A...6P}:
$\Omega_{\mathrm{m}} = 0.3111$, $\Omega_{\Lambda} = 0.6899$, $\Omega_{\mathrm{b}} = 0.0489$, $h = 0.6766$, and $\sigma_8 = 0.8102$.
All magnitudes are reported in the AB system \cite{1983ApJ...266..713O}.
We adopted a solar oxygen abundance of $12+\log(\mathrm{O/H}) = 8.69$ and a solar metallicity of $Z_\odot = 0.0142$ \cite{2009ARA&A..47..481A}.

\subsection*{JWST/MIRI MRS spectroscopy}

CEERS2-588 was observed with JWST/MIRI Medium Resolution Spectroscopy (MRS) \cite{2015PASP..127..646W,2023A&A...675A.111A}as part of a Cycle 3 program (\#4586; PI: Y. Harikane).
The Medium (B) and Short (A) gratings were used to cover \Oiii$\lambda5007$ and H$\alpha$ in Channels~1 and~2, respectively.
The total on-source exposure times were 9.5 hours for the Medium (B) grating and 17.8 hours for the Short (A) grating.
Dedicated background observations were obtained for both configurations.

The MRS observations were processed using version 1.20.0 of the JWST calibration pipeline and context 1462 of the Calibration Reference Data System (CRDS).
We followed the standard MRS pipeline procedure, with additional customized steps to improve the quality of the final calibrated products (see refs \cite{2023A&A...671A.105A,2024A&A...686A..85A,2025A&A...695A.250A} for details).
The final channel 1 and 2 cubes have spatial and spectral samplings of 0.13\arcsec\,$\times$\,0.13\arcsec\,$\times$\,0.8\,nm and 0.17\arcsec\,$\times$\,0.17\arcsec\,$\times$\,1.3\,nm, respectively \cite{2023AJ....166...45L}, and a resolving power of approximately 3500 \cite{2021A&A...656A..57L,2023MNRAS.523.2519J}.

From the reduced data cubes, one-dimensional spectra were extracted using a 0.\carcsec6-diameter circular aperture.
Because CEERS2-588 is unresolved in the MIRI images, aperture corrections were applied assuming the point-spread function.
Uncertainties were scaled to match the standard deviation measured over velocity ranges of $\pm20{,}000~\mathrm{km\,s^{-1}}$ from the expected line centers.

Neither H$\alpha$ nor [O\,\textsc{iii}]$\lambda5007$ was detected.
Upper limits on the line fluxes were derived assuming a line width of $200\ \mathrm{km\,s^{-1}}$.
The measured upper limits on line fluxes and rest-frame equivalent widths are listed in Extended Data Table~\ref{tab_spec}.

To verify that the non-detection of the {\Oiii} and H$\alpha$ emission lines is not caused by any pointing issues in the MIRI/MRS observations, we collapsed the MIRI/MRS datacubes along the wavelength axis to construct continuum images.
In the continuum images, a nearby source (R.A.$=$14:19:37.56, Decl.$=+$52:56:45.8) within the MRS field of view is clearly detected at its expected position.
This confirms that the MIRI/MRS observations were properly executed and that the spatial alignment is reliable.
Therefore, the non-detection of {\Oiii} and H$\alpha$ emission lines in CEERS2-588 is unlikely to be due to instrumental or pointing problems.

\subsection*{JWST/MIRI imaging}

In the same Cycle~3 program, MIRI imaging was obtained in the F560W and F770W filters, with total exposure times of 2.5 and 2.8 hours, respectively.
The F770W observations were taken simultaneously with the MRS background exposures, while the F560W image was taken as another observation.
Additional F1000W imaging targeting another $z=11$ galaxy in this field \cite{2022ApJ...940L..55F} was obtained during MRS observations for CEERS2-588, but this galaxy was not detected in this band.

The MIRI imaging data were reduced following refs \cite{2024ApJ...968....4P,2024ApJ...969L..10P} using the Rainbow pipeline, which is based on the version 1.19.41 of the JWST calibration pipeline and context 1413 of the Calibration Reference Data System (CRDS), and includes additional steps to handle the complex MIRI background.
A key component of the Rainbow workflow is the ``superbackground" strategy, in which a background template is constructed from other images in the same filter taken within a three-month period, and known sources are masked to avoid overestimating the background.
This procedure produces a highly uniform background in terms of both level and noise, and has been shown to improve the depth of the final mosaics by up to 0.8 mag compared to standard archive products.
Further details are provided in Appendix A of ref \cite{2024ApJ...968....4P}.

Photometry was performed using \textsc{SExtractor}\cite{1996A&AS..117..393B} (version~2.25.3) in the same manner as ref.~\cite{2023ApJS..265....5H}.
Fluxes were measured within a 0.\carcsec3-diameter circular aperture to minimise contamination from nearby sources, and aperture corrections were applied assuming the point-spread function.
The resulting flux densities are summarized in Extended Data Table~\ref{tab_phot}.
CEERS2-588 is detected in both F560W and F770W at $\sim6\sigma$ significance.

\subsection*{JWST/NIRCam and NIRSpec data}

\begin{figure}[t]
\centering
\includegraphics[width=0.99\textwidth, bb=9 6 922 492]{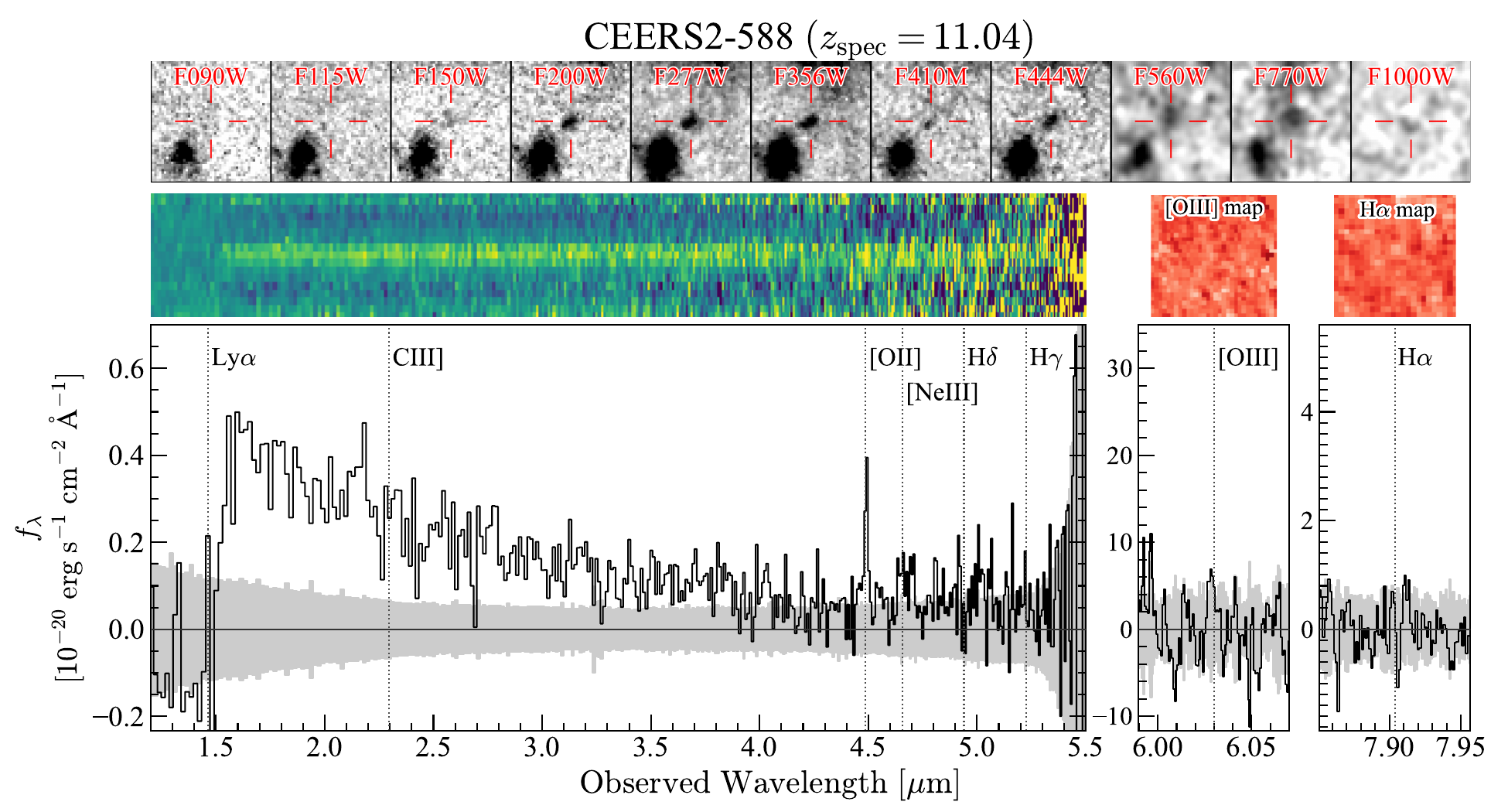}
\caption{{\bf JWST NIRCam, MIRI, and NIRSpec datasets of CEERS2-588.}
Top panels show 4\arcsec$\times$4\arcsec cutout images obtained with JWST NIRCam and MIRI. 
Middle panels present the NIRSpec two-dimensional spectrum and the MIRI/MRS maps at the expected wavelengths of \Oiii$\lambda5007$ and H$\alpha$. 
Bottom panels show the extracted one-dimensional NIRSpec and MIRI/MRS spectra (black), with the corresponding $1\sigma$ uncertainties (grey). 
The NIRSpec spectrum clearly reveals the \Oii$\lambda3727$ emission line, indicating the spectroscopic redshift of $z_\m{spec}=11.04$.
}
\label{fig_spec2}
\end{figure}

NIRCam and NIRSpec observations of CEERS2-588 were obtained from the CEERS program \cite{2025ApJ...983L...4F} (\#1345; PI: S.~Finkelstein) and program~\#2750 \cite{2023Natur.622..707A} (PI: P.~Arrabal Haro), respectively.
We used data reduced and publicly released through the DAWN JWST Archive.
Extended Data Fig.~\ref{fig_spec2} presents the NIRCam and NIRSpec data of CEERS2-588.

NIRCam photometry was measured using a 0.\carcsec3-diameter aperture following ref.~\cite{2023ApJS..265....5H} and is included in Extended Data Table~\ref{tab_phot}.
We corrected the NIRSpec spectrum for a slit loss by scaling it to match the NIRCam photometry, comparing the observed NIRCam magnitudes with pseudo-magnitudes calculated from the NIRSpec spectrum using the NIRCam filter response curves.
In the NIRSpec PRISM spectrum, the [O\,\textsc{ii}]$\lambda3727$ emission line is detected at a significance of $5.2\sigma$.
The rest-frame equivalent width of the {\Oii} line, $\sim100$ \AA, is comparable to typical galaxies at high redshift \cite{2024ApJ...976..193R}, disfavoring an extremely high ionizing-photon escape fraction.
A positive signal is present at the expected wavelength of [Ne\,\textsc{iii}]$\lambda3869$ at the $2.4\sigma$ level; however, this feature is conservatively treated as a non-detection and a $3\sigma$ upper limit is adopted.
These emission-line measurements are listed in Extended Data Table~\ref{tab_spec}.

\subsection*{Star formation rates}

Star formation rates were estimated from H$\alpha$ and UV luminosities.
Given the non-detection of H$\alpha$, the H$\alpha$-based star formation rate is constrained to be $<4.4~M_\odot~\mathrm{yr^{-1}}$, while the UV-based star formation rate is $8.2~M_\odot~\mathrm{yr^{-1}}$, using the calibrations in ref.~\cite{2024ApJ...977..133C}, assuming the Chabrier initial mass function \cite{2003PASP..115..763C}.
This difference is due to the different characteristic timescales traced with the H$\alpha$ and UV emission ($\sim5$ and $\sim100$ Myr, respectively).
Indeed, the ratio of the H$\alpha$- to UV-based star formation rate is $<0.54$, significantly lower than that observed in other galaxies at $z>10$ \cite{2025arXiv251219695H,2024ApJ...975..245C,2025A&A...695A.250A,2024ApJ...973...81H,2025ApJ...988L..10K}, suggesting a recent quenching event \cite{2025ApJ...994...14P}.
We did not use metal emission lines such as \Oii$\lambda3727$ to estimate the star formation rate, because their calibrations strongly depend on the physical conditions of the interstellar medium, including metallicity.
Instead, we incorporate the observed emission-line fluxes, including metal lines, into the SED fitting with varying the metallicity as a free parameter, to constrain the star formation history.

\subsection*{Gas-phase metallicity}

\begin{figure}[h]
\centering
\includegraphics[width=0.32\textwidth,bb=7 25 368 419]{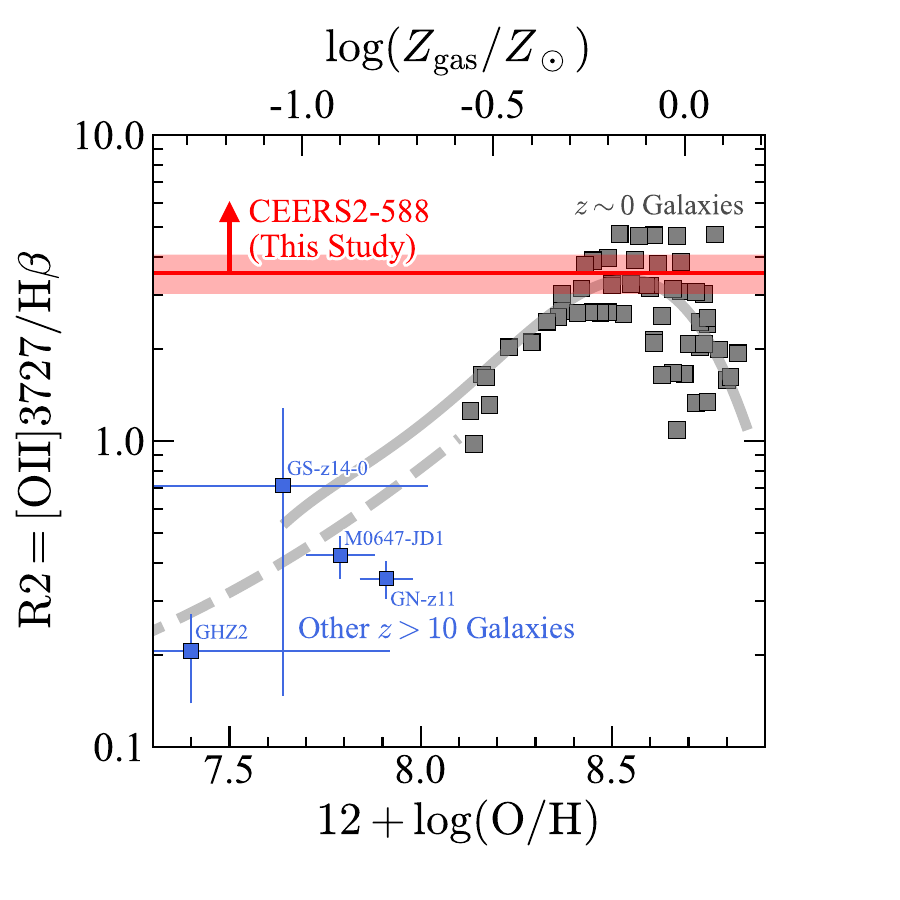}
\includegraphics[width=0.32\textwidth,bb=7 25 368 419]{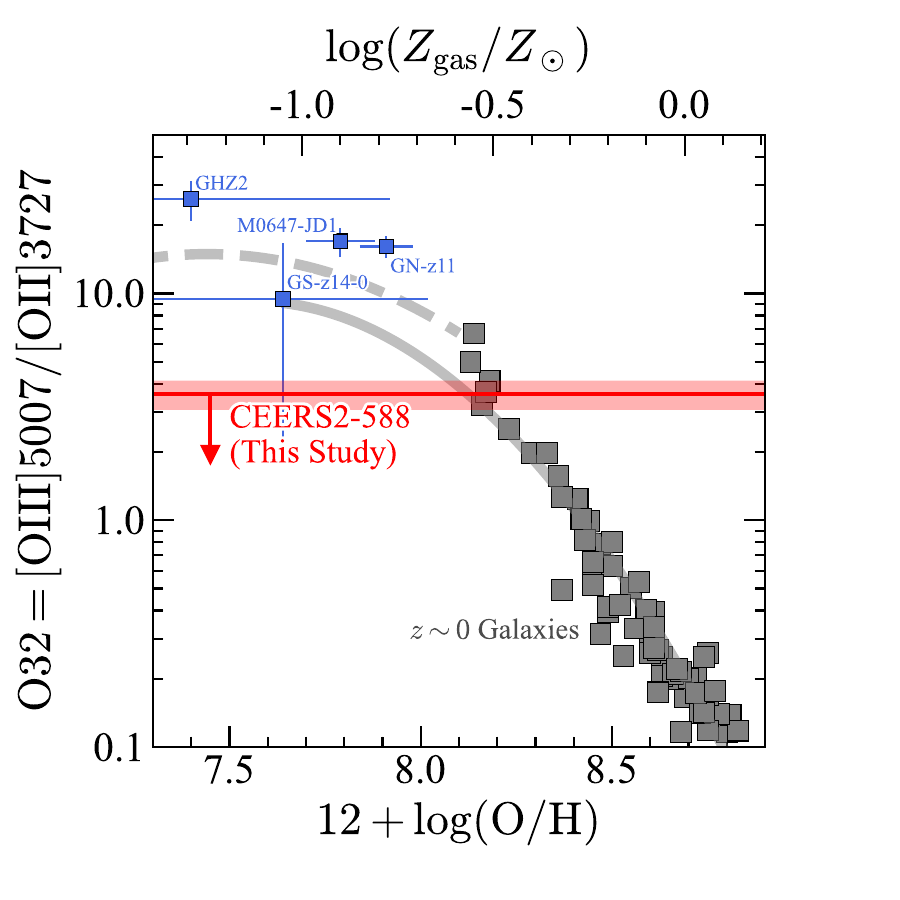}
\includegraphics[width=0.32\textwidth,bb=7 25 368 419]{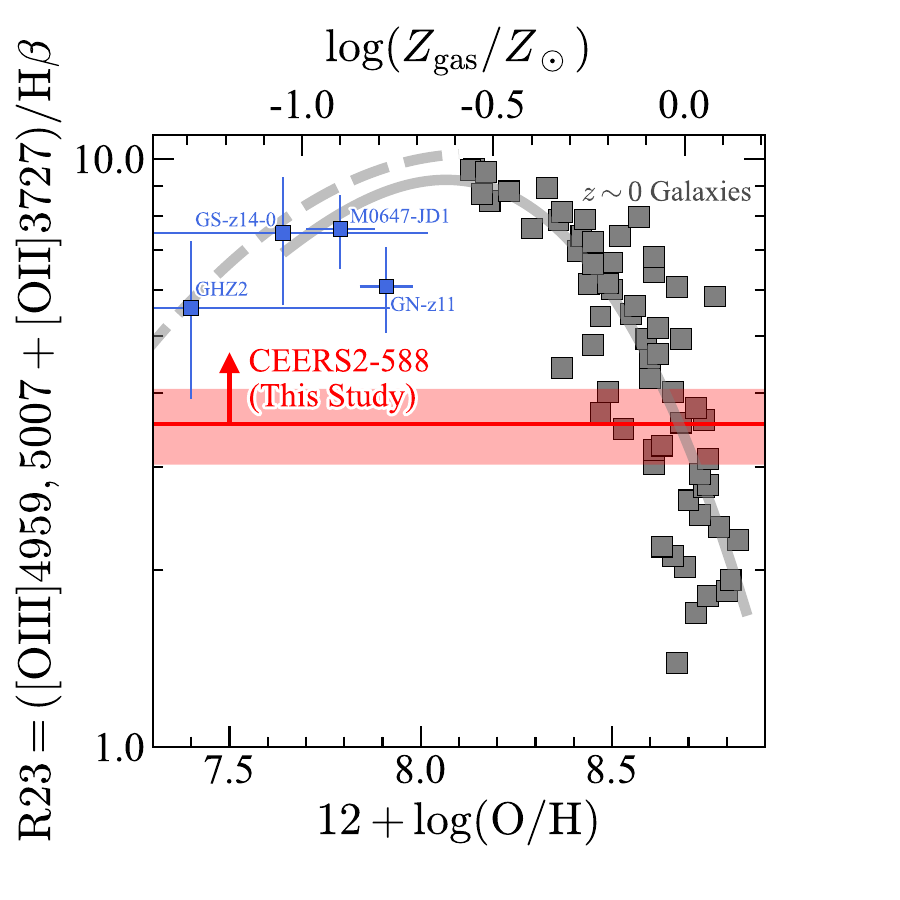}
\caption{\textbf{Constraints on the gas-phase metallicity of CEERS2-588.}
Each panel shows a strong-line ratio used as a metallicity diagnostic as a function of oxygen abundance: 
$\mathrm{R2}=$\Oii$\lambda3727$/H$\beta$ (left), $\mathrm{O32}=$\Oiii$\lambda5007$/\Oii$\lambda3727$ (middle), and $\mathrm{R23}=($\Oiii$\lambda4959,5007+$\Oii$\lambda3727)$/H$\beta$ (right). 
Red lines and shaded regions indicate the constraints on the line ratios of CEERS2-588 with $1\sigma$ uncertainties. 
Grey solid curves show empirical relations for massive galaxies \cite{2017MNRAS.465.1384C,2020MNRAS.491..944C}, while grey dashed curves correspond to relations for young galaxies at $z \sim 0$ \cite{2022ApJS..262....3N}. 
Grey squares represent stacked spectra of $z \sim 0$ SDSS galaxies \cite{2017MNRAS.465.1384C}. 
The combined strong-line constraints indicate a gas-phase metallicity of $12+\log(\mathrm{O/H}) \simeq 8.6$ for CEERS2-588, significantly higher than those measured for other galaxies at $z>10$ (blue squares) \cite{2025arXiv251219695H,2024ApJ...975..245C,2025A&A...695A.250A,2024ApJ...973...81H}.
}
\label{fig_Z}
\end{figure}

The gas-phase metallicity of CEERS2-588 was estimated using strong-line diagnostics.
Extended Data Fig.~\ref{fig_Z} compares the observed line ratios with empirical calibrations \cite{2017MNRAS.465.1384C,2020MNRAS.491..944C,2022ApJS..262....3N} and with stacked spectra of $z \sim 0$ SDSS galaxies \cite{2017MNRAS.465.1384C}.
From the measured [O\,\textsc{ii}]$\lambda3727$ flux and the upper limit on H$\beta$ inferred from H$\alpha$ assuming Case~B recombination, we obtained a constraint of $\mathrm{R2} > 3.5$.
This value is significantly higher than those measured for other galaxies at $z>10$ \cite{2025arXiv251219695H,2024ApJ...975..245C,2025A&A...695A.250A,2024ApJ...973...81H} and comparable to local massive galaxies.
We therefore obtained a metallicity of $12+\log(\mathrm{O/H}) = 8.55^{+0.10}_{-0.13}$, which we adopt as the fiducial value.
Dust extinction was assumed to be negligible based on the SED fitting; allowing for modest extinction would increase the inferred metallicity and does not affect our conclusions.
Constraints from the O32 and R23 diagnostics are consistent with the R2-based estimate.

\subsection*{SED fitting}

SED fitting was performed using \textsc{Bagpipes} version~1.3.2 \cite{2018MNRAS.480.4379C,2019MNRAS.490..417C}.
Free parameters included stellar mass, star formation history, metallicity, ionization parameter, and dust attenuation, while the redshift was fixed to the spectroscopic value.
The fitting incorporated NIRCam and MIRI photometry, the NIRSpec spectrum, and upper limits on the H$\alpha$ and [O\,\textsc{iii}]$\lambda5007$ line fluxes from the MIRI spectra.

We assumed the Kroupa initial mass function \cite{2001MNRAS.322..231K}, the Calzetti dust attenuation law \cite{2000ApJ...533..682C}, and the intergalactic medium attenuation model of ref.~\cite{2014MNRAS.442.1805I}.
The star formation history was parameterized using five age bins spanning 0--10, 10--25, 25--50, 50--130, and 130--300~Myr, with a continuity prior applied \cite{2019ApJ...876....3L}.
Flat priors were adopted for the remaining parameters over the ranges
$0 < E(B-V) < 1$~mag,
$-2.3 < \log(Z/Z_\odot) < 0.7$,
$-5 < \log U < 0$, and
$8 < \log(M_\ast/M_\odot) < 11$.
Sampling was performed using the \textsc{nautilus} nested sampler \cite{2023MNRAS.525.3181L}.

The best-fitting SED is shown in Fig.~\ref{fig_sed}, and the derived parameters are summarized in Extended Data Table~\ref{tab_target}.
Quantities based on a Kroupa initial mass function were converted to a Chabrier initial mass function by multiplying by 0.97.
The best-fitting model reproduces all photometric and spectroscopic constraints, and the inferred metallicity is consistent with the strong-line estimates.
The dust attenuation is negligible, indicating that the relatively red UV slope of CEERS2-588 ($\beta_{\rm UV}=-1.7$ \cite{2025ApJ...988...86Y}), compared to other galaxies at similar redshift \cite{2024MNRAS.531..997C,2024MNRAS.529.4087T,2025ApJ...995...43A}, is primarily due to its evolved stellar population rather than dust extinction.
We note that an SED of little red dots cannot reproduce the MIRI-band fluxes of CEERS2-588, because its optical continuum is much flatter than that of little red dots \cite{2025ApJ...995...21T}.

\subsection*{Ionizing photon production efficiency}

The ionizing photon production efficiency was calculated from the extinction-corrected H$\alpha$ and UV luminosities as
\begin{equation}
\xi_{\mathrm{ion}} = \frac{N(\mathrm{H^0})}{L_{\mathrm{UV}}},
\end{equation}
where $N(\mathrm{H^0})~[\mathrm{s^{-1}}] $ is the intrinsic ionizing photon production rate.
This quantity was estimated from the H$\alpha$ luminosity using the conversion of ref.~\cite{1995ApJS...96....9L},
\begin{eqnarray}
L_{\mathrm{H}\alpha}^{\mathrm{int}}~[\mathrm{erg\,s^{-1}}] &=& 1.36 \times 10^{-12} N_{\mathrm{obs}}(\mathrm{H^0})\notag\\
&=& 1.36 \times 10^{-12} (1 - f_{\mathrm{esc}}^{\mathrm{ion}}) N(\mathrm{H^0}),
\end{eqnarray}
where $f_{\mathrm{esc}}^{\mathrm{ion}}$ is the ionizing photon escape fraction.
We adopted the dust attenuation inferred from the SED fitting, $E(B-V)=0.04$, and assumed $f_{\mathrm{esc}}^{\mathrm{ion}} = 0$.
The resulting value of $\xi_{\mathrm{ion}}$ is reported in Extended Data Table~\ref{tab_target}.

\subsection*{Connection to other galaxies at $z>11$}

\begin{figure}[h]
\centering
\includegraphics[width=0.9\textwidth,bb=6 10 488 355]{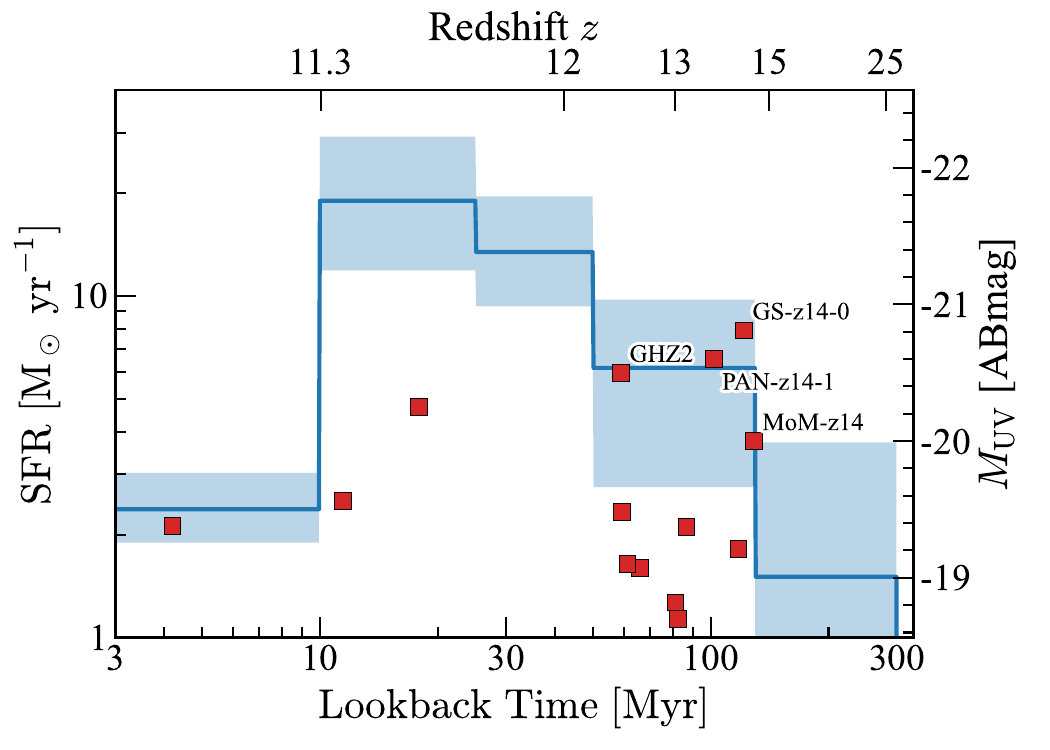}
\caption{\textbf{Connection of CEERS2-588 to other galaxies at $\mathbf{z>11}$.}
The blue line and shaded region show the star formation history of CEERS2-588 inferred from SED fitting.
Red squares indicate spectroscopically confirmed galaxies at $z>11$ \cite{2025arXiv250821708R,2025ApJ...995L..74B}.
}
\label{fig_sfh_gal}
\end{figure}

In Extended Data Fig.~\ref{fig_sfh_gal}, we compare the star formation history inferred from the SED fitting with the UV magnitudes of spectroscopically confirmed galaxies at $z>11$ \cite{2025arXiv250821708R,2025ApJ...995L..74B,2026arXiv260111515D}.
The predicted star formation rates of CEERS2-588 at $z\sim12-14$ are comparable to those of GHZ2 \cite{2024ApJ...972..143C,2025NatAs...9..155Z,2024ApJ...975..245C}, PAN-z14-1 \cite{2026arXiv260111515D}, and MoM-z14 \cite{2025arXiv250511263N}, suggesting that these galaxies could be progenitors of CEERS2-588.
The inferred star formation history extends to $z>15$, implying the presence of galaxies at even earlier epochs, consistent with recent reports of galaxy candidates at $z>15$ \cite{2025ApJ...991..179P,2025A&A...704A.158C,2025ApJ...992...63W,2026arXiv260115959H}.

\subsection*{Theoretical models}

In this study, we compared our results for CEERS2-588 with predictions from a variety of theoretical models, including FirstLight \cite{2023ApJ...953..140N}, BlueTides \cite{2017MNRAS.469.2517W}, FIRE-2 \cite{2018MNRAS.478.1694M,2024ApJ...967L..41M}, Santa Cruz-SAM \cite{2019MNRAS.490.2855Y}, UniverseMachine \cite{2019MNRAS.488.3143B,2020MNRAS.499.5702B}, FLARES \cite{2021MNRAS.500.2127L,2023MNRAS.518.3935W}, Astraeus \cite{2024A&A...686A.138C}, Millennium-TNG \cite{2023MNRAS.524.2594K}, and GAEA \cite{2025arXiv251103787C}.
In the right panel of Fig.~\ref{fig_Ms}, we show the metallicity predictions from FirstLight ($z=11$), FIRE-2 ($z=11$), FLARES ($z=10$), Astraeus ($z=10$), Millennium-TNG ($z=11$), and GAEA ($z=9.84$).
In the left panel of Fig.~\ref{fig_SFR}, we present the star formation rate predictions from FirstLight ($z=11$), BlueTides ($z=11$), FIRE-2 ($z=11$), UniverseMachine ($z=11.10$), FLARES ($z=10$), Santa Cruz-SAM ($z=10$), Millennium-TNG ($z=11$), and GAEA ($z=9.84$).

\newcounter{exttable}
\renewcommand{\thetable}{\arabic{exttable}}
\renewcommand{\tablename}{Extended Data Table}
\renewcommand{\theHtable}{ext.\arabic{extfigure}}
\setcounter{table}{0}

\stepcounter{exttable} 
\begin{table}[h]
\caption{Properties of CEERS2-588}
\begin{tabular}{@{}lcl@{}}
\toprule
Parameter & Value  & Description\\
\midrule
R.A. & 14:19:37.59 & Right Ascension (J2000)\\
Decl. & $+$52:56:43.8 & Declination (J2000)\\
$z_\mathrm{spec}$ & $11.04$ & Spectroscopic Redshift\cite{2024ApJ...960...56H}\\
$M_\mathrm{UV}$ [mag] & $-20.4$ & Absolute UV magnitude\cite{2024ApJ...960...56H}\\
$\beta_\mathrm{UV}$ & $-1.74\pm0.25$ & UV slope\cite{2025ApJ...988...86Y}\\
$r_\mathrm{e}$ [pc] & $453^{+120}_{-82}$ & Effective radius\cite{2025ApJ...991..222O}\\
$\mathrm{SFR_\mathrm{H\alpha}}$ [$M_\odot$ yr$^{-1}$] & $<4.4$ & Star formation rate from H$\alpha$\\
$\mathrm{SFR_\mathrm{UV}}$ [$M_\odot$ yr$^{-1}$] & $8.2$ & Star formation rate from UV\\
$\mathrm{log}\xi_\mathrm{ion}$ [Hz erg$^{-1}$] & $<24.9$ & Ionizing photon production efficiency\\
$12+\mathrm{log(O/H)}$ (Lines) & $8.55^{+0.10}_{-0.13}$ & Metallicity from strong line methods\\
$\mathrm{log}(M_*/M_\odot)$ & $9.05^{+0.07}_{-0.09}$ & Stellar mass from Bagpipes\\
$t_\mathrm{age}$ [Myr] & $82^{+33}_{-30}$ & Mass-weighted age from Bagpipes\\
$\mathrm{E(B-V)}$ [mag] & $0.04^{+0.03}_{-0.04}$ & Dust Attenuation from Bagpipes\\
$12+\mathrm{log(O/H)}$ (Bagpipes) & $8.59^{+0.18}_{-0.21}$ & Metallicity from Bagpipes\\
$\mathrm{log}U_\mathrm{ion}$ & $-3.73^{+0.60}_{-0.85}$ & Ionization parameter from Bagpipes\\
\botrule
\end{tabular}
\footnotetext{Errors are $1\sigma$ and upper limits are $3\sigma$.}
\label{tab_target}
\end{table}

\stepcounter{exttable} 
\begin{table}[h]
\caption{Photometric Measurements of CEERS2-588}
\begin{tabular}{@{}lccc@{}}
\toprule
Instrument & Observed Wavelength  & Flux Density & Reference\\
 & [$\mu$m]  & [nJy] & \\
\midrule
HST ACS/F435W & 0.43 & $<32.0$ & C25\cite{2025arXiv251008743C} \\
HST ACS/F606W & 0.60 & $<19.8$ & C25\cite{2025arXiv251008743C} \\
HST WFC3/F125W & 1.25 & $<42.7$ & C25\cite{2025arXiv251008743C} \\
HST WFC3/F160W & 1.54 & $<35.5$ & C25\cite{2025arXiv251008743C} \\
JWST NIRCam/F090W & 0.90 & $<14.6$ & This Study \\
JWST NIRCam/F115W & 1.16 & $<14.9$ & This Study \\
JWST NIRCam/F150W & 1.50 & $22.4\pm4.9$ & This Study \\
JWST NIRCam/F200W & 1.99 & $43.2\pm3.7$ & This Study \\
JWST NIRCam/F277W & 2.77 & $45.5\pm2.5$ & This Study \\
JWST NIRCam/F356W & 3.58 & $46.5\pm2.5$ & This Study \\
JWST NIRCam/F410M & 4.08 & $35.6\pm5.6$ & This Study \\
JWST NIRCam/F444W & 4.42 & $50.8\pm4.0$ & This Study \\
JWST MIRI/F560W & 5.65 & $65.4\pm10.7$ & This Study \\
JWST MIRI/F770W & 7.66 & $53.3\pm9.0$ & This Study \\
JWST MIRI/F1000W & 9.97 & $<110.1$ & This Study \\
\botrule
\end{tabular}
\footnotetext{Errors are $1\sigma$ and upper limits are $3\sigma$.}
\label{tab_phot}
\end{table}

\stepcounter{exttable} 
\begin{table}[h]
\caption{Emission Line Fluxes of CEERS2-588}
\begin{tabular}{@{}ccc@{}}
\toprule
Line & Line Flux  & Rest-Frame Equivalent Width\\
 & [$10^{-19}$ erg s$^{-1}$ cm$^{-2}$]  & [\AA] \\
\midrule
{[\sc{O\,ii}]}$\lambda$3727 &  $6.7\pm1.0$ &  $97.6\pm18.7$\\
{[Ne\,\sc{iii}]}$\lambda$3869 &  $<4.2$ &  $<61.3$\\
H$\delta$ &  $<3.6$ &  $<52.6$\\
H$\gamma$ &  $<3.9$ &  $<56.9$\\
{[\sc{O\,iii}]}$\lambda$5007 &  $<24.1$ &  $<421.6$\\
H$\alpha$ &  $<5.4$ &  $<177.8$\\
\botrule
\end{tabular}
\footnotetext{Errors are $1\sigma$ and upper limits are $3\sigma$.}
\label{tab_spec}
\end{table}

\stepcounter{extfigure} 

\stepcounter{extfigure} 

\stepcounter{extfigure} 

\backmatter

\bmhead{Acknowledgements}
We are grateful to Mirko Curti, Pratika Dayal, and Hiroto Yanagisawa for providing the data points of their papers, and Yoshihisa Asada, Claude-Andre Faucher-Giguere, Andrea Ferrara, Adriano Fontana, Chiaki Kobayashi, Julian Munoz, and Alessandro Trinca for useful comments and discussions.
This work is based on observations made with the NASA/ESA/CSA James Webb Space Telescope. The data were obtained from the Mikulski Archive for Space Telescopes at the Space Telescope Science Institute, which is operated by the Association of Universities for Research in Astronomy, Inc., under NASA contract NAS 5-03127 for JWST.
These observations are associated with programs ERS-\#1345, DDT-\#2750, and GO-\#4586. 
The authors acknowledge the JWST ERS-\#1345 and DDT-\#2750 teams led by Steven Finkelstein and Pablo Arrabal Haro, respectively, for developing their observing programs.
The NIRCam and NIRSpec data products presented herein were retrieved from the Dawn JWST Archive (DJA). DJA is an initiative of the Cosmic Dawn Center (DAWN), which is funded by the Danish National Research Foundation under grant DNRF140.
The MIRI data presented in this article were obtained from the Mikulski Archive for Space Telescopes (MAST) at the Space Telescope Science Institute. 
This publication is based upon work supported by the World Premier International Research Center Initiative (WPI Initiative), MEXT, Japan, the Japan Society for the Promotion of Science (JSPS) Grant-in-Aid for Scientific Research (24H00245, 25H00674), the JSPS Core-to-Core Program (JPJSCCA20210003), and the JSPS International Leading Research (22K21349).
This work was supported by the joint research program of the Institute for Cosmic Ray Research (ICRR), University of Tokyo.
YH acknowledges support from the Sumitomo Foundation, the Ito Science Promotion Society, and the Yamaguchi Scholarship Foundation.
PGP-G acknowledges support from grant PID2022-139567NB-I00 funded by Spanish Ministerio de Ciencia, Innovaci\'on y Universidades MCIU/AEI/10.13039/501100011033, FEDER {\it Una manera de hacer Europa}.
J.A.-M. acknowledges support by grants PID2024-158856NA-I00 \& PIB2021-127718NB-I00 from the Spanish Ministry of Science and Innovation/State Agency of Research MCIN/AEI/10.13039/501100011033 and by “ERDF A way of making Europe” and by grant CSIC/BILATERALES2025/BIJSP25022.

\bmhead{Code availability}
The MIRI data were processed with the JWST data reduction pipeline available at \url{https://jwst-docs.stsci.edu/jwst-science-calibration-pipeline}.
The NIRCam and MIRI photometric measurements were conducted with {\sc sextractor} available at \url{https://github.com/astromatic/sextractor}.

\bmhead{Data availability}

The JWST NIRCam and NIRSpec data used in this study are available at \url{https://dawn-cph.github.io/dja/}.
Other datasets generated and/or analyzed during this study are available from the corresponding author upon reasonable request.

%
%
%
%
%


\bibliography{reference}

\end{document}